\documentclass[twocolumn,runningheads]{svjour2}
\smartqed  
\usepackage{graphicx}
\journalname{Astrophysics and Space Science}
\begin{document}

\title{Nucleonic gamma-ray production in Pulsar Wind Nebulae 
}


\author{D. Horns, F. Aharonian, A.I.D. Hoffmann, A. Santangelo
}


\institute{D. Horns, A.I.D. Hoffmann, A. Santangelo  \at
Institute for Astronomy and Astrophysics\\
Karl-Eberhards University T\"ubingen\\
              Sand 1\\
	      72076 T\"ubingen \\
	      \and
	      F. Aharonian \at
	      Max-Planck-Institut f. Kernphysik Heidelberg\\
	      P.O. Box 10\,39\,80\\
	      69117 Heidelberg
}

\date{Received: date / Accepted: date}

\maketitle

\begin{abstract}
  Observations of the inner radian of the Galactic disk at very high energy
  (VHE) gamma-rays have revealed at least 16 new sources. Besides shell type
  super-nova remnants, pulsar wind nebulae (PWN) appear to be a dominant source
  population in the catalogue of VHE gamma-ray sources.  Except for the Crab
  nebula, the newly discovered PWN  are resolved at VHE gamma-rays to be
  spatially extended (5-20 pc). Currently, at least  3 middle aged ($t>10$ kyrs) PWN
  (Vela X, G18.0-0.7, and G313.3+0.6 in the ``Kookaburra'' region) and  1 young
  PWN MSH 15-5{\it2} ($t=1.55$~kyrs) have been identified to be VHE emitting
  PWN (sometimes called ``TeV Plerions'').  Two more candidate ``TeV Plerions''
  have been identifed and have been reported at this conference \cite{Svenja}.
  In this contribution, the gamma-ray emission from Vela X is explained by a
  nucleonic component in the pulsar wind. The measured broad band spectral
  energy distribution is compared with the expected X-ray emission
  from primary and secondary electrons. The observed
  X-ray emission and TeV emission from the three middle aged PWN are compared 
  with each other.
\end{abstract}

\section{Introduction}
\label{intro}
The discovery of extended gamma-ray emission from pulsar wind nebulae (PWN) opens an exciting
possibility to study the acceleration of particles in ultra-relativistic
shocks. Acceleration at relativistic shocks is of relevance for the
understanding of the mostly non-thermal emission seen from relativistic jets in
active galactic nuclei (AGN) and gamma-ray bursts (GRBs). While the Lorentz
factor assigned to the relativistic flow in AGN and GRBs is currently believed
to be in the range of $\Gamma\approx 10\ldots100$ based upon arguments of
opacity against pair production, the Lorentz factor of pulsar winds $\gamma$ is largely
undetermined. Using the potential drop of the last open field lines as a constraint:
$\gamma\le e\Phi_\mathrm{open}/(m_e c^2)=2\times 10^8 (\dot P_{-13}/P)^{1/2}$
with $P$ indicating the rotational period and $dP/dt=\dot P_{-13}\times
10^{-13}$~s/s the rate of slowing down. The Lorentz factor has been determined
in model dependent ways to be in the range of $10^6-10^7$ for the Crab nebula
(see e.g. \cite{Arons96} and \cite{BogoAhar00}). Even if this value is
possibly smaller in other PWN, we are witnessing particle acceleration in
ultrarelativistic shocks driven by the energetic pulsar wind.  The total energy
available by slowing down of the pulsar (assuming a moment of inertia $I=I_{45}
10^{45}$~g~cm$^2$): $\dot E=10^{34}~I_{45} \dot P_{-13}/P^3$~erg/s.\\
Theoretical calculations of Fermi-type acceleration in relativistic shocks
using widely different approaches (see e.g.  \cite{BedOstro98}, \cite{Achter01}) lead to the consistent conclusion, that (largely
independent of the detailed parameters of the shock)  particles in the
downstream plasma are accelerated to follow a universal power-law for large
Lorentz factors of $s= 2.2\ldots 2.3$ which is compatible with e.g. 
the observed X-ray synchrotron spectrum of e.g. the Crab nebula.\\
However, recent progress in the understanding of the downstream turbulence
spectrum, a revision of this ``universal acceleration'' picture may be
necessary (see e.g. \cite{LePeRe06}, \cite{NiemOs06}).
 In the more recent calculations, the
particle spectrum is found to be softer  than in previous calculations (see
above), to deviate from a universal power-law, and  to show a stronger dependence on the Lorentz factor of the upstream
medium and on the particular implementation of downstream turbulence. Similar
conclusions have been found independently with particle-in-cell (PIC) simulations
(\cite{Hoshi92}, \cite{Spit05}, \cite{AmAr06}).\\
The presence of ions in the pulsar wind obviously complicate the structure of
the shock and have been described e.g. in \cite{HoshiAr91},
 \cite{Gall92}, \cite{Hoshi92}.
 Hoshino et al. (1992) \cite{Hoshi92}  found that an admixture of ions in the wind
can lead to acceleration of positrons in the downstream region by resonant
absorption of magnetosonic waves emitted by the gyrating ions. In a recent 
PIC simulation \cite{AmAr06}
a larger ratio of ion mass $m_i$ to electron mass $m_e$ in the simulation has been used 
(previously, PIC simulations had been limited to values of $m_i/m_e\approx 20$) and
various energy fraction of ions have been considered. The overall efficiency of 
acceleration has been found to increase with the relative energy fraction carried by ions in the wind. 
The authors also show that the spectrum of non-thermal particles varies.
\\
 The presence of the so-called  wisps in the Crab nebula has been used to argue
 for the presence of ions in the wind (see e.g. \cite{GallAr96}): the
 compression caused by the ions leads to increased magnetic field and
 correspondingly more intense synchrotron radiation. Wisp-like features have
 been observed from PSR B1509-58 \cite{Gaens02}, however, the predicted
 time variation for the northern arc was not found  in later observations
 \cite{Del06}. The
 Vela PWN shows variable features along its jet which have been interpreted to
 be the result of a kink instability \cite{Pav03}. The ring like features in the Vela PWN
 have not been seen to vary with time even though the expected time scale
 should be similar to the one observed from the Crab nebula.   \\ 
 Besides the existence of wisp-like structures near the wind shock,
 ions are expected to leave other observable signatures. The downstream energy
 distribution  of the ions follows a relativistic Maxwellian distribution with
 some modifications due to the energy loss of ion energy transfered to the
 non-thermal tail of accelerated pairs \cite{Hoshi92}. The
 temperature of the distribution is close to $\gamma m_i c^2$. At some distance
 to the shock, the ions will move diffusively outwards, loosing energy by
 adiabatic expansion. Gamma-rays and neutrinos will be produced predominantly
 in inelastic scattering of the nucleons on the ambient medium. 
 Depending on the diffusion coefficient in the PWN, a large
 fraction of the particle energy can be converted into gamma-rays and
 neutrinos. \\ 
 Here, we consider specifically the Vela X PWN, which has
 recently been detected to emit VHE gamma-rays \cite{HESS06}. In the
 final section, we will discuss similarities and differences to the other ``TeV
 Plerions''. For a more general review of gamma-ray production in PWN see
 \cite{Bed06} in these proceedings.
 \begin{figure}
\centering
  \includegraphics[width=\linewidth]{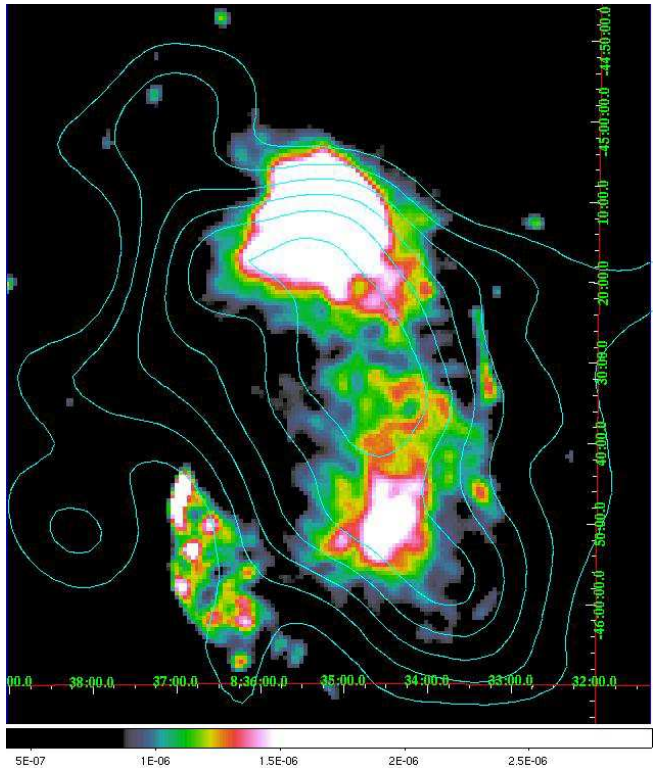}
\caption{In grey scale (see the online proceedings for a color version): ASCA 2-10 keV mosaic showing the flux in bins of 0.25 arc~min$^2$ size. The contours 
represent the HESS significance map \cite{HESS06} with the contours starting at $5~\sigma$ above the
background noise and incrementing by one standard deviation.}
\label{fig:1}       
\end{figure}

 \section{Vela pulsar and Vela X} 
 The Vela X region was initially discovered as an extended radio emitting
 region \cite{Ris58}, south of the Vela Pulsar PSR~B0833-45 located in the
 center of the Vela supernova remnant.  The distance to the Vela pulsar is well
 known from parallax measurements to be 290 pc \cite{Car01}.  The Vela Pulsar
 spins with a period of 89.3 ms and slows down with $\dot P=1.247\times 10^{-13}$~s/s which gives a current spin down luminosity of $\dot E=7\times
 10^{36}$~erg/s and an age of $t\approx $11~kyrs \cite{Tay93}. \\ X-ray
 emission from the Vela X region was discovered with the ROSAT X-ray telescope
 \cite{MarOegel95} and initially assumed to be a ``jet'' from the pulsar. The
 spectral range of the ROSAT PSPC instrument was not sufficient to clearly
 identify a non-thermal tail in the observed energy spectrum. Later high
 resolution measurements of the Vela pulsar with Chandra revealed a compact
 X-ray nebula with a double torus structure \cite{Helf01}. The morphology of the Vela
 compact nebula and its orientation suggests that the Vela X region is not a ``jet'' but rather the
 result of the interaction of the middle aged pulsar wind system with the
 reverse shock of the supernova shock wave. Hydrodynamic simulations of the
 interaction of an expanding PWN inside the shell remnant indicate that for
 middle aged PWN, the reverse shock will start to interact with the PWN leading
 ultimately to a compression of the PWN \cite{Blond01}. For the specific case
 of asymmetries in the density of the ambient medium into which the external
 shock of the supernova blastwave expands, the reverse shock will start to
 interact earlier with the expanding PWN along the direction where the ambient
 medium is more dense. The Vela X PWN shows indeed such asymmetry which could
 be the result of the interaction of the evolved PWN with an asymmetric reverse
 shock.

 \section{Non-thermal X-rays from Vela X: New results from ASCA}
 The X-ray observations of the Vela X region with ROSAT are not conclusive on
 the existence of a non-thermal power-law component in the energy spectrumThe ASCA satellite with its four X-ray telescopes \cite{Serle95}
 equipped with two Gas Imaging Spectrometers (GIS, \cite{Oha96})
 and two Solid State Imaging Spectrographs (SIS, \cite{Bur94}) is ideally suited
 to image an extended region like Vela X at energies between 2--10~keV.
 Given the size of the Vela X region, we do not consider the SIS data with 
 the smaller field of view. The GIS
 data were screened following the standard screening criteria. For the 4 early pointings
 (50021000, 50021010, 50021020, 50021030) no
 rise-time selection (Ohashi et al. 1996) was possible.  These pointings cover mainly the region
 at the northern end of Vela X.
 See also Table~1 for an overview of the observation number, observation date, and exposure (combined GIS2 \& GIS3). 
%
\begin{table}[t]
\caption{ASCA pointings used to generate the mosaic shown in Fig.~1}
\centering
\label{tab:1}       
\begin{tabular}{lll}
\hline\noalign{\smallskip}
ID & Obs. date &  exposure  \\
   &           & [ksec]     \\[3pt]
\tableheadseprule\noalign{\smallskip}
23043000 & 04/15/1995 & 34  \\
23043010 & 04/15/1995 & 34 \\
25038000 & 12/02/1997 & 76 \\
50021000 & 05/12/1993 & 20 \\
50021010 & 06/26/1993 & 20 \\
50021020 & 07/14/1993 & 26 \\
50021030 & 10/08/1993 & 30 \\
\noalign{\smallskip}\hline
\end{tabular}
\end{table}
 Since we are interested in the morphology of the Vela X region in non-thermal
 X-rays , we consider for the image only 
 events with energies exceeding 2 keV. The data from the two GIS detectors have
 been added to increase the statistics. The particle background was estimated
 from Earth night sky observations and subtracted off the skymap and the
 resulting excess map with 0.5 arc min bins was divided by the exposure map. The analysis is similar
 to the one described in \cite{Rob01}.
 \\
 The resulting flux image is shown in Fig.~\ref{fig:1} in grey scale. Note,
 the surface brightness is rather faint on the level of $5\times 10^{-6}$
 counts/(s~cm$^2$~arc~min$^2$).  The overlaid contours are from the excess map
 of the VHE gamma-ray source \cite{HESS06}.  \\
 The X-ray morphology observed in the ASCA image is very similar to the ROSAT
 picture \cite{MarOegel95} and shows bright emission centered on the position
 of the pulsar with a narrow extension to the south and a re-brightening at
 the southern end of Vela X.  A bright feature appears to the south east of
 the Vela X region which is marginally significant and unfortunately located
 at the edge of the field of view. More observations of this region at X-ray
 energies could be of interest as there appears to be an increase of the VHE
 signal towards that same unexplored region.  \\
 A more quantitative study of the X-ray morphology at different energies is
 beyond the scope of this paper. However it is noteworthy, that the size of
 the X-ray emitting region at energies of 2--10~keV appears to be smaller than
 the observed size in the soft energy band as e.g. the ROSAT PSPC data
 indicate.  \\
 In order to investigate possible spectral variations, we have sub-divided the
 extension of Vela X in three regions excluding the bright feature related to
 the compact nebula in the north by excising a 6 arc~min radius region
 centered on the Vela Pulsar which includes the compact X-ray nebula
 \cite{Man05}.  The energy spectra of all three regions are compatible with a
 mixture of a thermal component and a power-law with a 
 photon index of 2 and varying fluxes. The southern tip has also been observed
 with XMM-Newton. The ASCA energy spectrum of the southern tip has been cross-checked with the
 XMM-Newton spectra which show good agreement as both observations 
 are well fit by a power-law component to be present up to 8~keV. This
 is consistent with the result of a combined ROSAT and ASCA SIS analysis \cite{MaOeg97} 
 of the southern tip region.
 \\
 In order to compare the observed X-ray emission and the VHE emission, an energy spectrum
 covering the entire length of the X-ray emitting region excluding a 6 arc min 
 region centered on the Vela pulsar has been extracted. For the background estimate, 
 the dim region to the east of the Vela X has been used. The resulting energy spectrum 
 is well fit by a power-law with photon index $\approx2$ and is shown in Fig.~\ref{fig:3} together
 with the ROSAT flux and the BeppoSAX/XMM-Newton combined analysis of the compact nebula emission \cite{Man05}.
\begin{figure}
\centering
  \includegraphics[width=\linewidth]{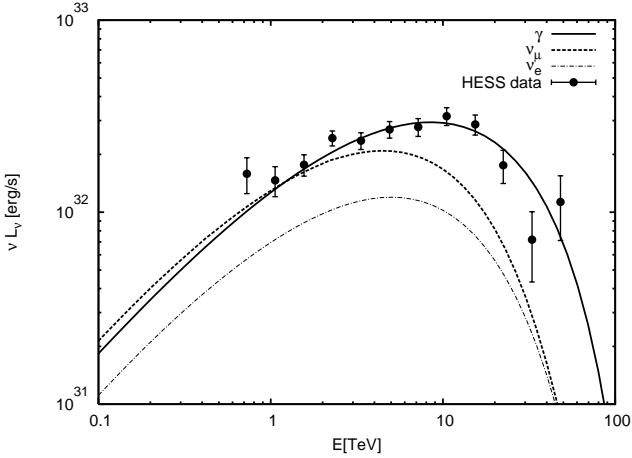}
\caption{A comparison of the measured VHE gamma-ray spectrum  \cite{HESS06} with the model fit. The dashed 
and dot-dashed curves show the expected neutrino spectra.}
\label{fig:2}       
\end{figure}
%
\begin{figure}
\centering
  \includegraphics[width=1.0\linewidth]{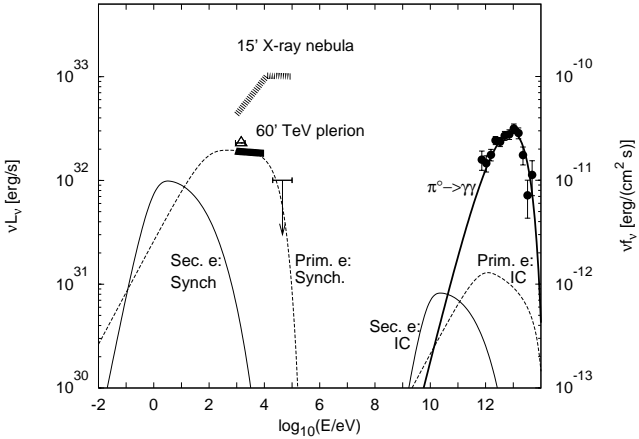}
\caption{Broad band spectral energy distribution from the Vela X region 
including the combined XMM-Newton/BeppoSAX result on the compact nebula centered on the Vela Pulsar (``15' X-ray nebula''), the ROSAT result (open triangle),
the ASCA spectrum excluding the compact nebula (``60' TeV plerion''),
and the corresponding INTEGRAL upper limit from the same region (99~\% c.l.)}
\label{fig:3}       
\end{figure}
\section{Nucleonic gamma-rays from Vela X}
 \subsection{Observations}
 The first clear indication of a gamma-ray signal from the Vela X region was
 found with the H.E.S.S. telescopes with a luminosity in the oberved energy
 range $L_{0.6-65~\mathrm{TeV}}\approx 10^{33}$~erg/s \cite{HESS06} at the
 distance of 290~pc. The spatial extension of the TeV plerion is slightly
 larger than the size of the X-ray emitting region and extends for a full width
 at half maximum (FWHM) of 5.7~pc along the major axis and for a FWHM of 4.3~pc
 along the minor axis (again at a distance of 290~pc).  The observed energy
 spectrum can be described by a power-law with a cut-off $dN/dE=N_0
 (E/1~\mathrm{TeV})^{-\Gamma}\cdot \exp(-E/E_c)$ with the best fit parameters
 $\Gamma=1.45\pm0.09_\mathrm{stat}\pm 0.2_\mathrm{sys}$ and
 $E_c=13.8\pm2.3_\mathrm{stat}\pm4.1_\mathrm{sys}$~TeV.  
 \subsection{Model}
 The observed gamma-ray spectrum can be
 modeled under the assumption that the pulsar wind carries a large fraction of
 the spin down energy in the form of nucleons that expand diffusively to fill
 the PWN. The pulsar releases most of its energy during the first few hundred
 years, when the spin down proceeds most rapidly. While the pulsar spins down,
 $\Phi_\mathrm{open}$ decreases and therefore it is natural to assume that the
 corresponding Lorentz factor of the particle wind decreases as well, even
 though we do not have as yet direct evidence for this evolutionary effect.\\
 Given the short time in which the particles are injected, it is reasonable to
 assume that we have an instantaneous injection event and therefore, the
 particle distribution is assumed to follow a relativistic Maxwellian
 distribution with a temperature $k_BT=\gamma m_i c^2$. 
 \\ 
 When considering
 the relevant time-scales for the evolution of the particle distribution in
 comparison with the age of the Vela Pulsar, we assume Bohm-type diffusion and
 energy loss for the nucleons in inelastic scattering with the ambient medium
 with a density of $n=0.6~\mathrm{cm}^{-3}$. This a lower limit to the overall
 density of this region derived from the emission measure of the plasma in
 X-rays. In principle, the total target density could be higher than this value
 in case of the presence of cold gas. The diffusion time
 $t_\mathrm{diff}=7300~\mathrm{yrs} Z^{-1} \eta (R_4)^2 B_{-5} E_{100}^{-1}$ is shorter than the age of the system and therefore,
 escape losses are negligible. The magnetic field of $10~\mu$G is a reasonable 
 value for the Vela X region (see also next section). 
 More detailed calculation \cite{horns06} indicate that the
 total energy in protons required to match the observed gamma-ray flux is
 $W_p\approx 10^{49}$~erg.
 \\ 
 A fit of the energy spectrum of gamma-rays from
 proton-proton interaction assuming a relativistic Maxwellian distribution is
 shown in Fig.~\ref{fig:2}. The $\pi^0$ decay spectrum has been calculated
 using the parametrization given in  \cite{keln06}.  The observed energy spectrum is
 well described for $E=80~\mathrm{TeV}$ as the energy of the Maxwellian which
 translates into $\gamma\approx 8\times 10^4$. 
 \\ 
 A characteristic signature
 of the nucleonic production mechanism is the production of neutrinos. Using
 the corresponding parametrization for the production of muon and electron
 flavor neutrinos \cite{keln06}, we calculate the differential energy spectra
 for the two flavors also shown in Fig.~\ref{fig:2}. 
 \\
 Given the high energy cut-off and the reasonable angular extension of the Vela X
 region, we consider Vela X among the best candidates for a detection as a neutrino
 source with the future Neutrino telescope in the mediterranean sea.
\subsection{X-ray emission from secondary electrons}
Due to the interaction of nuclei with ambient medium, neutral and charged
mesons are produced. While the neutral mesons (primarily $\pi^0$) decay to
produce $\gamma$-rays, charged pions decay into electrons and positrons. These
secondary electrons then in turn radiate mainly via synchrotron, inverse
Compton, and Bremsstrahlung. The energy loss time of the electrons depends on
the energy density of the background radiation field (magnetic field strength
and soft photon density) and the particle number density. In the case of Vela
X, synchrotron cooling dominates very likely with a cooling time expressed as a
function of the characteristic energy of synchrotron radiation emitted:
$t_{1/2}=1.2~\mathrm{kyrs} B_{-5}^{-3/2}E_{keV}^{-1/2}$.  The
magnetic field strength in the Vela X region is difficult to estimate. The
equipartition field derived from radio measurements of narrow structures can be
estimated to be $20-50~\mu$G. However, when taking the X-ray flux in the
considered region, the equipartition field is $B_\mathrm{eq}=4~\mu$G. For such 
a small magnetic field however, electrons are not efficiently confined in the Vela X region 
and should have escaped already and would fill a larger volume which in turn
should produce a considerably larger TeV plerion. Therefore, a magnetic field around $10~\mu$G is considered to be
a realistic parameter to match the observed size of the TeV plerion.\\
The synchrotron emission from secondary electrons is calculated taking into account radiative cooling
but neglecting escape losses (see above).  The resulting synchrotron spectrum assuming a $10~\mu$G magnetic
field after 11\,000 yrs is shown in Fig.\ref{fig:3} together with the relevant measurements. Clearly, 
the secondary electrons' contribution to the overall spectral energy distribution is negligible. In principle, 
the synchrotron component would dominate in the optical. However, the sensitivity for extended emission
from the Vela X region in the optical is at least one order of magnitude above the expectation \cite{mign03}. 
\subsection{X-ray emission from primary electrons}
In the framework of ion resonant acceleration, pairs are accelerated to follow
a power-law which extends up to a maximum energy given by the energy of the
upstream ions. With the observed gamma-ray spectrum, the upstream ion energy is
constrained to be of the order of 100~TeV. Given the energy losses of the downstream ions, the upstream value ($\gamma m_ic^2$)
is higher than the temperature of the Maxwellian downstream distribution which is subsequently modified by adiabatic losses.
 We assume for simplicity that the 
pulsar is currently injecting a wind with a Lorentz factor of $\approx 10^5$ infered from the 
fit of the observed gamma-ray spectrum neglecting time-dependent effects.  \\ The
primary electrons accelerated at the shock will therefore reach a maximum
energy of $E_\mathrm{max}\approx 100~A/Z~\mathrm{TeV}$ with $Z$ the charge and $A$ the mass
number of the nuclei in the wind \cite{Hoshi92}. Given the magnetic field in
the downstream region (see below), this value limits the maximum energy of emitted
synchrotron radiation.\\ The magnetic field at the shock can be calculated in the
standard MHD picture of PWN (e.g. \cite{KC84}) by considering the distance of
the shock to the pulsar, the total extent, and  luminosity of the nebula.  The
most recent estimates range from $B(\theta_s=33'')=72~\mu\mathrm{G}$
\cite{SefaOki03} assuming the position of the shock at an angular separation of
$33$ arc~sec while using the value of $21$~arc sec obtained by an elaborate
fitting method of the torus \cite{Ng04} increases
$B(\theta_s=21'')=113~\mu\mathrm{G}$. This value is obtained assuming a
magnetization of $\sigma=0.1$ which appears consistent with MHD simulations
modeling the size of the Vela nebula \cite{bogo05}. Considering the shock
compression the downstream magnetic field can reach values of
$B_\mathrm{max}\approx 2~ B(\theta_s)\approx 200~\mu\mathrm{G}$.\\
Correspondingly, the emitted synchrotron radiation of electrons is given by
$\epsilon_\mathrm{max}=50~\mathrm{keV}
(B_{-4})(A/Z)^2E_{100}^2$.\\
The detection of unpulsed X-ray emission up to 200~keV with INTEGRAL from the
Vela pulsar/PWN \cite{Herms} indicates that possibly the
nucleons would have to be only partially ionised to increase (A/Z) and
therefore the maximum energy of the pairs. Partially ionised nuclei have already been suggested in
\cite{Hoshi92} in order to avoid exceeding the Goldreich-Julien current. It is
quite interesting to note that most of the pulsars powering the TeV plerions have been also detected
as INTEGRAL hard X-ray sources (see Hoffmann et al. these proceedings). More
data on hard X-ray emission from TeV plerions is of great importance to
understand the in situ acceleration at the pulsar wind shock and its maximum
energy.\\ The Vela X region has not been detected to emitt X-rays beyond
$\approx 8$~keV.  Taking all available data with INTEGRAL results in a
conservative upper limit on the energy flux in the 20--60~keV band as shown in
Fig.~\ref{fig:3}. Comparing the upper limit with the spectrum obtained with
ASCA it is evident, that a cut-off in the X-ray spectrum with an energy
$\approx 10$~keV is required not to violate the upper limit.\\ The origin of
the X-rays is very likely synchrotron emission from the primary electrons that
are accelerated at the shock and show a radiative cooling break which moves
from a few keV to below keV energies when extracting the energy spectra at
increasing angular separation from the Vela pulsar \cite{Man05}. This is
consistent when considering the Vela X region, where it turns out to be below keV
energies. Under the assumption of a constant injection rate with a power-law
with index $2$, the synchrotron spectrum for a $10~\mu$G field after 11\,000
yrs is shown in Fig.\ref{fig:3}. The total injected energy in electrons amounts
to $W_e\approx 10^{45}~B_5^{-2}$~erg between
$E_\mathrm{min}=0.01$~TeV and $E_\mathrm{max}=200~$TeV with the minimum energy chosen
close to $\gamma m_e c^2=0.5$~TeV. 
\section{Other TeV plerions} 
Besides the Vela X TeV plerion, at least two more middle aged PWN have been observed to be TeV plerions: G18.0-0.7
\cite{HESS05} and G313.3+0.6 \cite{HESSKooka} (the northern wing of the ``Kookaburra''). It is interesting to point out similarities and distinct
differences of these objects:
All three objects are spatially resolved to be extended at VHE gamma-rays. While the Vela X region measures $\approx 5$~pc across,
G18.0-0.7 (25.3~pc at 4~kpc distance) and G313.3+0.6 
(12.1~pc at 5.6~kpc distance) fill a substantially larger volume. The differences in size can be at least partially attributed
to the age: the Vela pulsar ($t=11~$kyrs) is nominally the youngest of the three objects (G18.0-0.7 is 21.5~kyrs and
G313.3+0.6 13~kyrs old). However, there are indications from the very small braking index of the Vela pulsar that 
the actual age of Vela is closer to 50~kyrs which would be more consistent with the rather cold neutron star surface.
\\
When comparing the spin down power of the pulsar with the radiative power
(either in X-rays or gamma-rays), it is interesting to note that while in
Vela, the radiative power accounts for less than 1 per cent of the spin down
power, it accounts for a few per cent in the other objects.  Finally, we
consider the ratio of the published extension (\cite{gaens03} for G18.0-0.7
and \cite{Ng05} for G313.3+0.6) aof the X-ray nebula
$r_\mathrm{X}$ to the radius of the TeV plerion $r_\mathrm{TeV}$:  for Vela X,
this ratio is fairly close to 1 while  it is only
$r_\mathrm{X}/r_\mathrm{TeV}\approx 0.1$ for G18.0-0.7 and G313.3+0.6. It is
beyond the scope of this paper to investigate the deeper reason for this large
difference, but it is clear that the spatially resolved gamma-ray spectroscopy
with H.E.S.S. combined with deep X-ray observations covering the same angular
region will greatly improve the understanding of the underlying physical
parameters.




\end{document}